\documentclass[usegraphicx,usenatbib]{mn2e}

\usepackage{graphicx}
\usepackage{natbib}
\usepackage{fixltx2e}
\usepackage{mathptmx}
\usepackage{amsmath}
\usepackage{wasysym}
\usepackage{url}
\usepackage{times}
\usepackage{epsfig}
\usepackage{ulem}

\newcommand{\Mpc}{\rm\; Mpc}
\newcommand{\kpc}{\rm\; kpc}

\newcommand{\km}{\rm\; km}

\newcommand{\cm}{\rm\; cm}

\newcommand{\yr}{\rm\; yr}

\newcommand{\s}{\rm\; s}

\newcommand{\ks}{\rm\; ks}

\newcommand{\Ms}{\rm\; Ms}

\newcommand{\Msun}{\hbox{$\rm\thinspace M_{\odot}$}}

\newcommand{\Msunpyr}{\hbox{$\Msun\yr^{-1}\,$}}

\newcommand{\keV}{\rm\; keV}

\newcommand{\erg}{\rm\; erg}

\newcommand{\ergpcmsqps}{\hbox{$\erg\cm^{-2}\s^{-1}\,$}}

\newcommand{\ergps}{\hbox{$\erg\s^{-1}\,$}}

\newcommand{\keVps}{\hbox{$\keV\s^{-1}\,$}}

\newcommand{\kmps}{\hbox{$\km\s^{-1}\,$}}

\newcommand{\Zsun}{\hbox{$\thinspace \mathrm{Z}_{\odot}$}}

\newcommand{\amin}{\rm\; arcmin}

\newcommand{\asec}{\rm\; arcsec}

\newcommand{\kpcpasec}{\hbox{$\kpc\asec^{-1}\,$}}

\newcommand{\psqcm}{\hbox{$\cm^{-2}\,$}}
\newcommand{\pcmsq}{\hbox{$\cm^{-2}\,$}}
\newcommand{\pcmcu}{\hbox{$\cm^{-3}\,$}}

\begin{document}

\title[Inside the Bondi radius of M87]{Inside the Bondi radius of M87}
\author[H.R. Russell et al.]  
    {\parbox[]{7.in}{H.~R. Russell$^{1,2}$\thanks{E-mail: 
          hrr27@ast.cam.ac.uk}, A.~C. Fabian$^1$, B.~R. McNamara$^{2,3,4}$ and A.~E. Broderick$^{2,3}$ \\ 
    \footnotesize 
    $^1$ Institute of Astronomy, Madingley Road, Cambridge CB3 0HA\\
    $^2$ Department of Physics and Astronomy, University of Waterloo, Waterloo, ON N2L 3G1, Canada\\
    $^3$ Perimeter Institute for Theoretical Physics, Waterloo, Canada\\ 
    $^4$ Harvard-Smithsonian Center for Astrophysics, 60 Garden Street, Cambridge, MA 02138, USA\\
  }
}

\maketitle

\begin{abstract}
  \textit{Chandra} X-ray observations of the nearby brightest cluster
  galaxy M87 resolve the hot gas structure across the Bondi accretion
  radius of the central supermassive black hole, a measurement
  possible in only a handful of systems but complicated by the bright
  nucleus and jet emission.  By stacking only short frame-time
  observations to limit pileup, and after subtracting the nuclear PSF,
  we analysed the X-ray gas properties within the Bondi radius
    at $0.12-0.22\kpc$ ($1.5-2.8\asec$), depending on the black hole
    mass.  Within $2\kpc$ radius, we detect two significant
  temperature components, which are consistent with constant values of
  $2\keV$ and $0.9\keV$ down to $0.15\kpc$ radius.  No evidence was
  found for the expected temperature increase within $\sim0.25\kpc$
  due to the influence of the SMBH.  Within the Bondi radius, the
  density profile is consistent with $\rho{\propto}r^{-1}$.  The lack
  of a temperature increase inside the Bondi radius suggests that the
  hot gas structure is not dictated by the SMBH's potential and,
  together with the shallow density profile, shows that the classical
  Bondi rate may not reflect the accretion rate onto the SMBH.  If
  this density profile extends in towards the SMBH, the mass accretion
  rate onto the SMBH could be at least two orders of magnitude less
  than the Bondi rate, which agrees with Faraday rotation measurements
  for M87.  We discuss the evidence for outflow from the hot gas and
  the cold gas disk and for cold feedback, where gas cooling rapidly
  from the hot atmosphere could feed the cirumnuclear disk and fuel
  the SMBH.  At $0.2\kpc$ radius, the cooler X-ray temperature component
    represents $\sim20\%$ of the total X-ray gas mass and, by losing
    angular momentum to the hot gas component, could provide a fuel
    source of cold clouds within the Bondi radius.
\end{abstract}



\section{Introduction}
\label{sec:intro}


Accretion onto the supermassive black hole (SMBH) at the centre of the
giant elliptical galaxy M87 is powering relativistic jet activity that
heats the surrounding hot cluster atmosphere and stifles galaxy growth
(\citealt{FormanM8705,FormanM8707}).  Deep \textit{Chandra} X-ray
observations of the intracluster medium have found large cavities
filled with radio emission where the X-ray gas has been displaced by
the expansion of the jet (\citealt{Bohringer95}; \citealt{Bicknell96}; \citealt{Young02};
\citealt{FormanM8705,FormanM8707}).  This energy input is likely suppressing the
rapid cooling of the dense, low temperature X-ray gas at the centre of
the cluster, which would otherwise flood the brightest cluster galaxy
(BCG) with cold molecular gas and ensuing star formation
(eg. \citealt{PetersonFabian06}).  In general, this feedback from the
central AGN is thought to be the essential mechanism in models of
galaxy formation that truncates galaxy growth by restricting gas
cooling and star formation (\citealt{Croton06}; \citealt{Bower06};
\citealt{Hopkins06}).

\begin{table*}
\begin{minipage}{\textwidth}
\caption{Details of the \textit{Chandra} observations used for this analysis and best-fit nuclear spectral model parameters.}
\begin{center}
\begin{tabular}{l c c c c c c}
\hline
Obs. ID & Date & Exposure & Frame time & $n_{\mathrm{H}}$ & $\Gamma$ & Flux ($2-10\keV$)\\
 & & (ks) & (s) & ($10^{22}\psqcm$) & & ($10^{-12}\ergpcmsqps$) \\
\hline
352 & 29/07/2000 & 29.4 & 3.2 & - & - & - \\
1808 & 30/07/2000 & 12.8 & 0.4 & $0.08\pm0.01$ & $2.37\pm0.07$ & $0.70\pm0.04$ \\
11513 & 13/04/2010 & 4.7 & 0.4 & $0.06\pm0.02$ & $2.35\pm0.08$ & $1.42^{+0.10}_{-0.09}$ \\
11514 & 15/04/2010 & 4.5 & 0.4 & $0.05\pm0.02$ & $2.15\pm0.09$ & $1.4\pm0.1$ \\
11515 & 17/04/2010 & 4.7 & 0.4 & $0.04\pm0.02$ & $2.21\pm0.08$ & $1.6\pm0.1$ \\
11516 & 20/04/2010 & 4.7 & 0.4 & $0.02\pm0.02$ & $2.02\pm0.08$ & $1.7\pm0.1$ \\
11517 & 05/05/2010 & 4.7 & 0.4 & $0.08\pm0.02$ & $2.34\pm0.08$ & $1.54^{+0.10}_{-0.09}$ \\
11518 & 09/05/2010 & 4.0 & 0.4 & $0.06\pm0.02$ & $2.3\pm0.1$ & $1.17^{+0.10}_{-0.09}$ \\
11519 & 11/05/2010 & 4.7 & 0.4 & $0.08\pm0.02$ & $2.4\pm0.1$ & $1.02\pm0.08$ \\
11520 & 14/05/2010 & 4.6 & 0.4 & $0.11\pm0.02$ & $2.5\pm0.1$ & $1.01^{+0.08}_{-0.07}$ \\
13964 & 05/12/2011 & 4.5 & 0.4 & $0.07\pm0.02$ & $2.31^{+0.10}_{-0.09}$ & $1.17^{+0.09}_{-0.08}$ \\
13965 & 25/02/2012 & 4.6 & 0.4 & $0.04\pm0.02$ & $2.11\pm0.09$ & $1.5\pm0.1$ \\
14973 & 12/03/2013 & 4.4 & 0.4 & $0.05\pm0.02$ & $2.3\pm0.1$ & $1.18\pm0.09$ \\
14974 & 12/12/2012 & 4.6 & 0.4 & $0.05\pm0.03$ & $2.2\pm0.1$ & $1.17\pm0.09$ \\
Stack & & 54.7 & 0.4 & $0.057\pm0.006$ & $2.25\pm0.03$ & $1.30\pm0.03$ \\
\hline
\end{tabular}
\end{center}
\label{tab:obs}
\end{minipage}
\end{table*}

The central AGN is located in the densest region of the cluster's hot
atmosphere and therefore classical Bondi accretion of the hot gas is
often considered as a fuelling mechanism for feedback
(\citealt{Bondi52}; \citealt{Baganoff03}; \citealt{DiMatteo03}).  The gas will be accreted by the SMBH if it lies
within the Bondi radius, where the gravitational potential of the SMBH
dominates the thermal energy of the hot atmosphere.  The Bondi
radius is given by $r_{\mathrm{B}}=2GM_{\mathrm{BH}}/c_{s}^2$, where
$M_{\mathrm{BH}}$ is the mass of the black hole and $c_s$ is the sound
speed of the hot gas.  Using \textit{Chandra} observations of M87,
\citet{DiMatteo03} found that the Bondi accretion rate is able to
provide a sufficient supply of fuel to power the AGN.  However, Bondi
accretion of the hot gas may be insufficient to fuel the most powerful
radio jet outbursts (eg. \citealt{Rafferty06}; \citealt{McNamara11}).  Cold gas accretion
from circumnuclear disks of atomic and molecular gas may be required to supplement
the fuel supply (eg. \citealt{PizzolatoSoker05}).  

For accretion at the Bondi rate, the radiative output from the nucleus
in M87 is orders of magnitude below the expectation from standard thin
disk accretion (\citealt{Fabian88}; \citealt{Fabian95};
\citealt{DiMatteo00}).  The black hole at the centre of the Milky Way,
Sgr A$^*$, is a good example of a low luminosity AGN with a radiative
output several orders of magnitude below that implied by the Bondi
accretion rate and the standard $10\%$ radiative efficiency
(eg. \citealt{Baganoff03}; \citealt{Genzel10}).  This observed
radiative inefficiency of the accretion disk is consistent with an
advection-dominated accretion flow where the gas is unable to radiate
the gravitational potential energy locally and the trapped energy is
instead advected inward (\citealt{Ichimaru77}; \citealt{Rees82};
\citealt{Narayan94}; \citealt{Narayan08}).  Alternatively, the mass
inflow at the Bondi radius may not reach the event horizon of the SMBH
but is instead ejected from the black hole's sphere of influence
(ADIOS; \citealt{Blandford99,Blandford04}) or circulates in convective
eddies (CDAF; \citealt{Narayan00}; \citealt{Quataert00CDAF}).

It has proven difficult to distinguish between these possibilities, or
indeed to determine the fuelling mechanism of the AGN, because it is
only possible to resolve the hot gas inside the Bondi radius in a
handful of systems.  Deep \textit{Chandra} observations of two
  of the best targets, Sgr A$^*$ and NGC3115 (\citealt{Baganoff03};
  \citealt{Wong11}), are consistent with flat density profiles with
  $\rho{\propto}r^{-1}$ inside the Bondi radius (\citealt{Wang13};
  \citealt{Wong14}).  This is consistent with recent numerical
simulations which show outflows in the form of winds off the accretion
flow preventing all but a fraction of the Bondi rate from reaching the
SMBH (eg. \citealt{Yuan12}; \citealt{Li13}).  However, whilst such a
low mass accretion rate is adequate for these quiescent systems, it is
insufficient to fuel powerful radio-jet outbursts without requiring
efficiencies much greater than 100\% (\citealt{Nemmen14}).  Such high
efficiencies are theoretically achievable if the black hole spin
energy can be extracted (\citealt{Tchekhovskoy11,Tchekhovskoy12};
\citealt{McKinney12}) but alternatively, the inflow rate may be
significantly increased by cold gas accretion, which compensates for
the outflowing mass (\citealt{PizzolatoSoker05}; \citealt{Gaspari13}).

At a distance of only $\sim16\Mpc$, M87 at the centre of the Virgo
cluster is one of few systems for which \textit{Chandra} can resolve
the Bondi radius at $\sim2\asec$ ($\sim0.15\kpc$;
\citealt{DiMatteo03}; \citealt{Garcia10}).  Hosting a powerful radio
jet outburst and a circumnuclear gas disk (\citealt{Ford94};
\citealt{Harms94}; \citealt{Macchetto97}), M87 provides an excellent
comparison system to NGC3115 and Sgr A$^*$ to determine the properties
of the putative accretion flow.  However, the bright nuclear point
source and jet knots make it challenging to extract the properties of
the underlying cluster gas across the Bondi radius.  Early
\textit{Chandra} observations of M87 in 2000 showed that the bright
nucleus and jet emission would not be piled up in a $0.4\s$ frame time
observation (\citealt{Wilson02}; \citealt{Perlman05}).
\citet{DiMatteo03} used this short $12.8\ks$ observation to determine
the gas properties using single temperature spectral models in to
$2\asec$ radius.  In 2005, major flaring of the jet knot HST-1,
located only $0.85\asec$ from the nucleus, then resulted in strong
pileup even in short frame-time observations (\citealt{Harris06}) and
prevented follow up studies.  However, HST-1 has slowly declined in
brightness since the peak in 2005 and \citet{Harris11} found that by
2010 the intensity had dropped to the level observed in 2000.

Here we stack twelve $5\ks$ monitoring observations of M87 taken since
April 2010 (PI Harris) and subtract the bright nuclear emission to
determine the density profile inside the Bondi radius and study the
temperature structure on these scales.  Section
\ref{sec:dataanalysis} describes the observations, data reduction and
subtraction of the nuclear emission.  In section \ref{sec:results}, we
determine the temperature and density structure within the Bondi
radius and calculate the cooling and dynamical time profiles.  We
discuss accretion of the hot gas atmosphere by the SMBH and the
evidence for hot and cold gas outflows in section \ref{sec:discussion}
before presenting our conclusions in section \ref{sec:conclusions}.
We assume a distance to M87 of $16.1\Mpc$ (\citealt{Tonry01}) to be
consistent with earlier analyses of the \textit{Chandra} datasets.
This gives a linear scale of $0.078\kpcpasec$ and $\sim2-5\times10^{5}$
gravitational radii per arcsec, depending on the black hole mass (section \ref{sec:Bondi}).  All errors are $1\sigma$ unless
otherwise noted.


\section{\textit{Chandra} data analysis}
\label{sec:dataanalysis}


\begin{figure*}
\begin{minipage}{\textwidth}
\centering
\includegraphics[width=0.45\columnwidth]{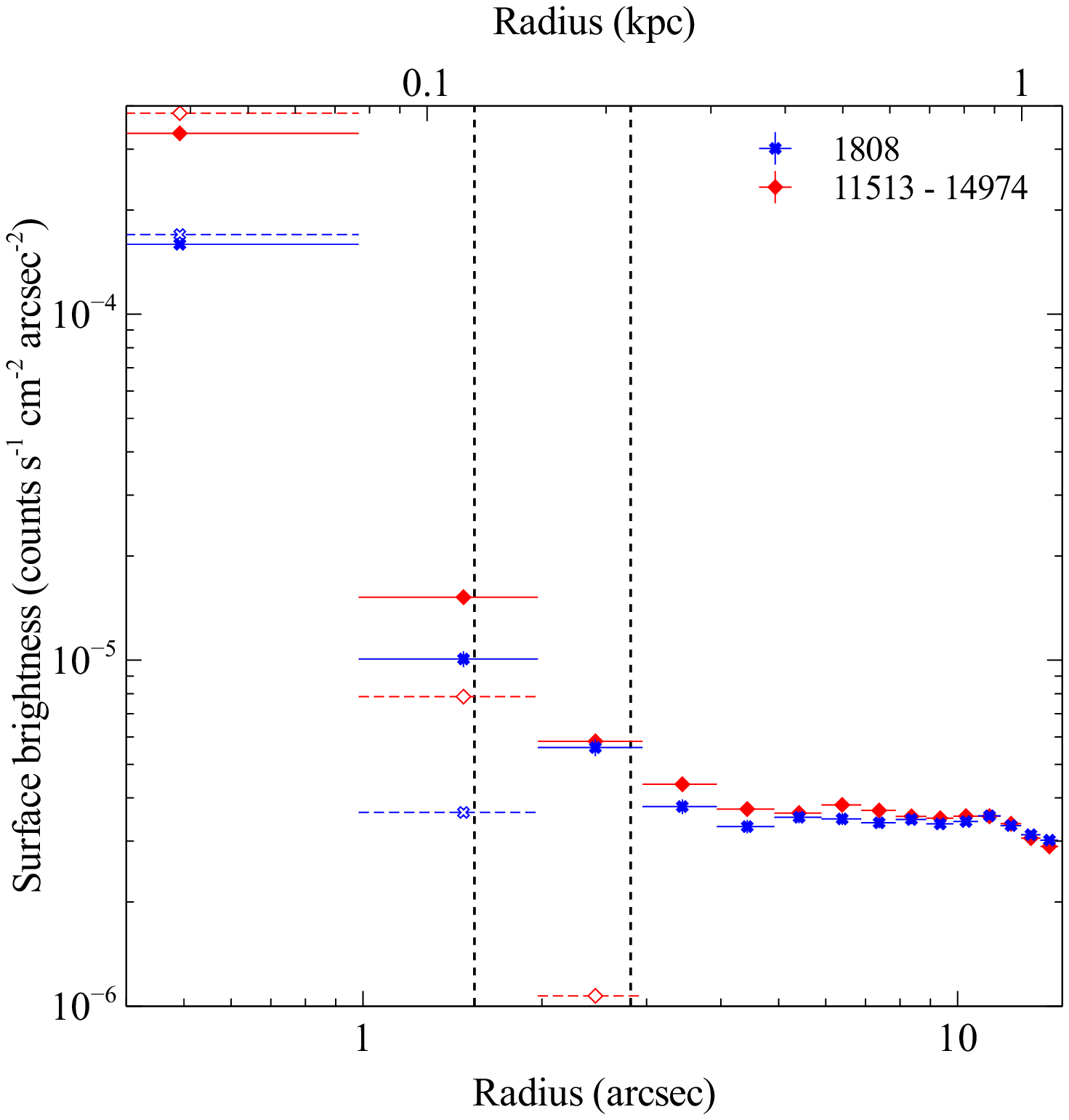}
\includegraphics[width=0.48\columnwidth]{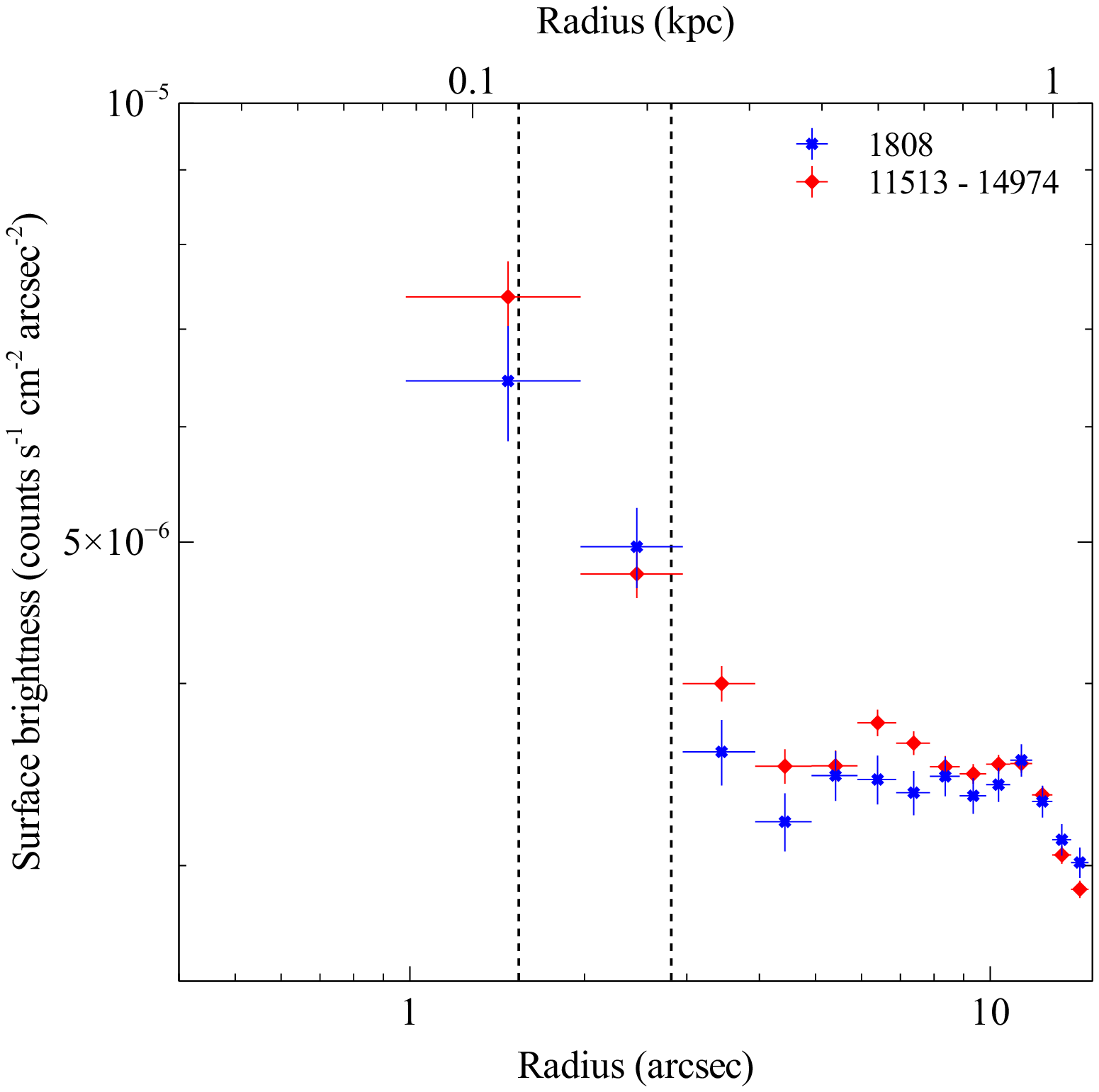}
\caption{Left: Background-subtracted surface brightness profiles in the energy range $0.5-7.0\keV$ for obs. ID 1808 and the summed obs. IDs 11513 to 14974.  The ChaRT simulations of the nuclear flux for each observation are shown by the open symbols.  Right: Same as left but with the nuclear flux subtracted from each profile.  The radial range for the Bondi radius is shown by the vertical dashed lines.}
\label{fig:psfprepostsub}
\end{minipage}
\end{figure*}

The centre of M87 has been observed regularly with \textit{Chandra}
since 2002 to monitor jet activity, understand the emission mechanism
and determine the site of the TeV flaring
(\citealt{Harris03,Harris06,Harris09,Harris11}).  For this study of
the central cluster gas properties, we have selected all archival
\textit{Chandra} observations of M87 taken since the intensity of
HST-1 dropped back to its 2000 level and after the TeV flaring event
in early April 2010 (\citealt{Ong10}; \citealt{Harris11}).  These
ACIS-S datasets were also compared with two of the ACIS-S
observations taken in 2000 to test the subtraction of the nuclear
emission (obs. ID 1808) and determine the gas properties to large
radii for the deprojection analysis (obs. ID 352).  We have not
included the short, $<1\ks$, exploratory observation, which
was taken to determine the nuclear count rate and therefore the frame
time suitable for the later observations (obs. ID 351).

All datasets were analysed with \textsc{ciao} version 4.6 and
\textsc{caldb} version 4.6.2 supplied by the \textit{Chandra} X-ray
Center (\citealt{Fruscione06}).  Level 1 event files were reprocessed
with the latest gain and charge transfer inefficiency correction
applied and then filtered to remove cosmic ray events.  Background
light curves were extracted from regions free of point sources and
used to remove periods affected by flares from the event files.
The combined dataset was also examined for any periods with high
  background count rates.  No flares were found in the short
frame-time observations and only two short flares occurred during
obs. ID 352.  The final cleaned exposure times are given in Table
\ref{tab:obs}.

The cluster emission from M87 extends across the entire
\textit{Chandra} field of view, therefore blank-sky background
observations were used to subtract the background from images and
spectra.  The appropriate blank-sky background dataset was processed
identically to the event file, reprojected to the same sky position
and normalized so that the count rate matched that of the event file
for the $9.5-12\keV$ energy band.  Point sources were identified using
a hard band ($3-5\keV$) image for each observation and excluded from
the analysis.  The shallow datasets from 2010 to 2013 were also
reprojected to a common position and combined to ensure that no faint
point sources were missed.  This method identified and excluded all
point sources found by \citet{Jordan04}.  The remaining contaminating
flux from unresolved low mass X-ray binaries and the stellar
population (cataclysmic variables and coronally active binaries) is
insignificant at only $\sim1\%$ of the total emission in M87
  in the $0.5-2\keV$ energy band (\citealt{Revnivtsev08}).  The
  cluster emission dominates at energies below $10\keV$ and the point
  source emission only becomes significant at higher energies.  The
emission from the jet was also carefully excluded.

\subsection{Subtracting the nuclear PSF}


Within $1\asec$, the nuclear point source is more than an order of
magnitude brighter than the cluster emission.  Although the cluster
emission dominates beyond $\sim3\asec$, the Bondi sphere is located
within this region.  It was therefore necessary to accurately model
and subtract the nuclear point spread function (PSF) to determine the
gas properties across the Bondi radius.  This analysis is described in
detail in the appendix, including an evaluation of the low-level
pileup, the nuclear spectral model and simulations of the PSF.  In
summary, following \citet{Russell10} and \citet{Siemiginowska10}, we
used the best-fit power law model for the nuclear spectrum (Table
\ref{tab:obs}) and the \textit{Chandra} ray-tracer (ChaRT;
\citealt{Carter03}) to simulate the nuclear PSF (photons scattered by
the \textit{Chandra} mirrors).  This simulation was then projected
onto the ACIS detector with \textsc{marx} and used to subtract the PSF
contamination from the cluster emission at each radius within the
cluster core.  We assume that all emission within $1\asec$ is
  from the AGN.  However the gas emission is likely to be $\sim5\%$ of
  the total, depending on the slope of the gas density profile (section A2).


The simulation was tested by comparing the PSF-subtracted surface
brightness profiles for two separate epochs (obs. ID 1808 and the
stacked 11513 to 14974 observations).  Fig. \ref{fig:psfprepostsub}
(left) shows the total emission profiles and the predicted
contribution of the PSF, which has varied significantly in flux
between the two epochs.  Fig. \ref{fig:psfprepostsub} (right) shows
the subtracted profiles, which incorporate the uncertainty on the
nuclear flux (Table \ref{tab:obs}).  The flux of the PSF simulation
was increased to compensate for the reduction due to mild pileup
within $1\asec$, which causes an oversubtraction of the AGN emission
in this region (section A3).  The subtracted profiles agree within the
errors showing that the majority of the variable nuclear emission has
been removed and only the cluster emission remains.  Although the
ChaRT/\textsc{marx} simulation is not expected to provide an exact
prediction of the PSF to large radii (eg. \citealt{Gaetz05}), the
nuclear flux is only significant to $\sim3\asec$ radius and the
excellent agreement between the subtracted profiles in
Fig. \ref{fig:psfprepostsub} (right) suggests this method is
sufficiently accurate for our analysis.  

The combined effect of the minor inaccuracies in the nuclear
  PSF subtraction due to the cluster contamination, mild pileup,
  nuclear spectral model and effective area reduction (section A3) was
  determined by comparing the subtracted profiles from these different
  epochs.  In Fig. \ref{fig:psfprepostsub}, the cluster surface
  brightness in the $1-2\asec$ radial bin, where the uncertainty due
  to the PSF subtraction is largest, agrees within the errors.  This
  suggests that the difference in flux between the two epochs is less than $\sim15\%$.
  The nuclear PSF flux is significantly lower in obs. ID 1808
  therefore it is more likely that the PSF contribution has been
  underestimated in the summed observations.  To be conservative, we
  have estimated the impact on the measured gas properties if the PSF
  flux is systematically higher by $20\%$.  Variation in the photon
  index by $\pm0.2$ and absorption by $\pm0.05\times10^{22}\pcmsq$
  produced uncertainties in the measured temperature and density
  values less than the statistical error and therefore is not
  significant in comparison to the uncertainty in the nuclear flux.

\subsection{Spectral analysis and deprojection}
\label{sec:clusterspec}

\begin{figure*}
\begin{minipage}{\textwidth}
\centering
\includegraphics[width=0.45\columnwidth]{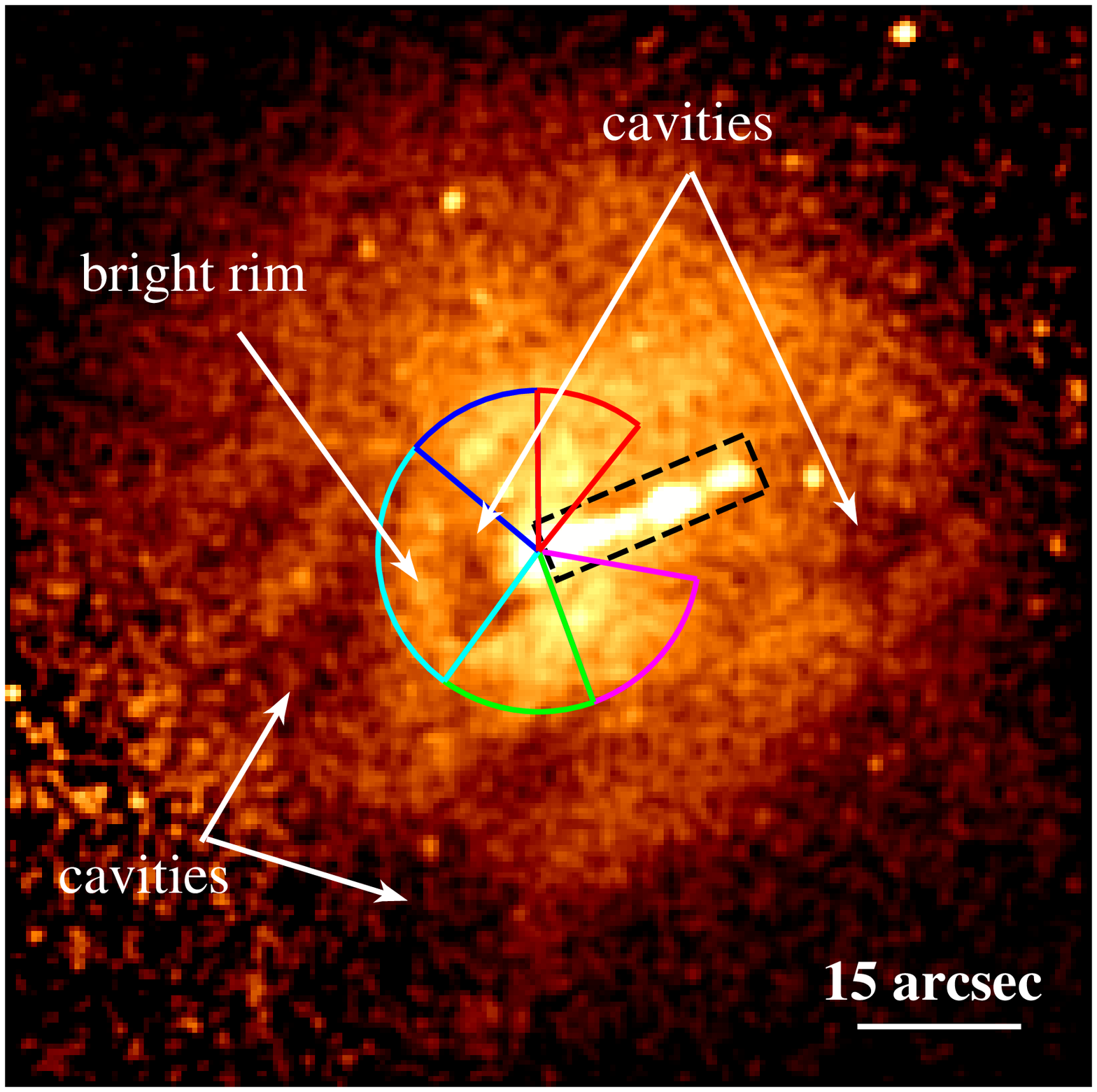}
\includegraphics[width=0.45\columnwidth]{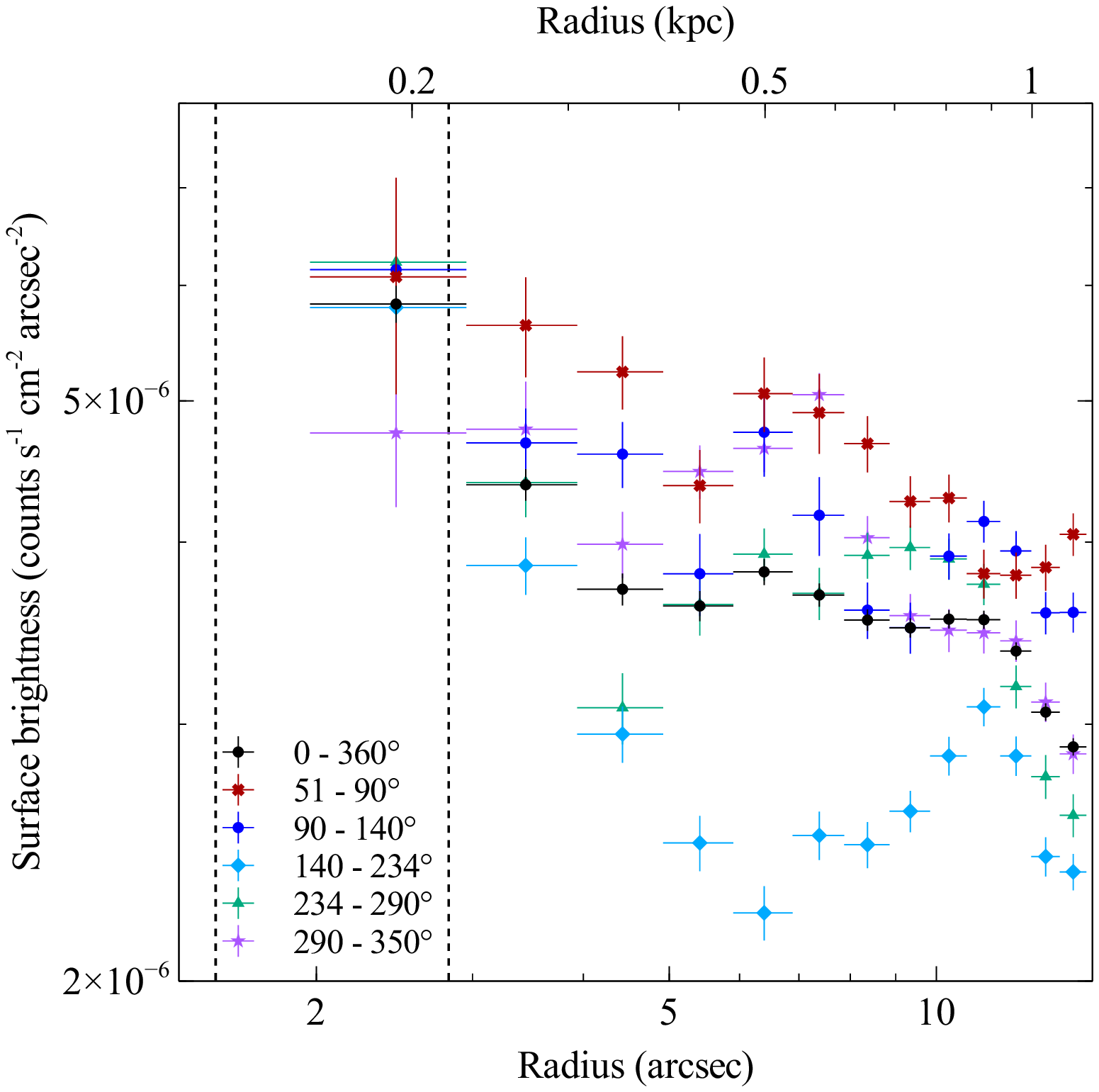}
\caption{Left: Exposure-corrected image showing the sectors used to
  compare the angular variation in the cluster surface brightness around the nucleus.  The excluded jet region is shown by the dashed black box.
  Right: Background-subtracted surface brightness profiles in the
  energy band $0.5-7.0\keV$ extracted from these sectors.  There is a cavity in the
  sector from $140-234^{\circ}$ which complicates the deprojection
  analysis.  The radial range for the Bondi radius is shown by the
  vertical dashed lines.}
\label{fig:angvariation}
\end{minipage}
\end{figure*}

Fig. \ref{fig:angvariation} (left) shows the complex cavity structures
and bright blobs and filaments of cool gas located at the centre of
M87.  These gas structures have been studied in detail by
\citet{Young02}, \citet{FormanM8705}, \citet{FormanM8707},
\citet{Million10} and \citet{Werner10}.  Clear depressions are
observed in the X-ray emission to the west of the nucleus beyond the
jet, immediately south-east of the nucleus and to the south and east
of that ($\sim20\asec$ from the nucleus).  These depressions are
cavities carved out by the radio jet and buoyantly rising through the
surrounding atmosphere (eg. \citealt{Churazov00};
\citealt{McNamara00}; \citealt{FabianPer00};
\citealt{McNamaraNulsen07}).  Fig. \ref{fig:angvariation} (right)
shows the angular variation in surface brightness around the nucleus.
The inner cavity to the south-east can be clearly seen in the
$140-234^{\circ}$ sector.  This cavity is surrounded by a rim of cool
gas that is visible as a bright ridge of emission.  The X-ray surface
brightness to the north and south of the nucleus is clumpy and several
point sources are visible, which were excluded from the profiles.
However, there is less angular variation in the surface brightness
inside $\sim4\asec$ radius.  We have therefore proceeded with
azimuthally-averaged density and temperature profiles.

Spectra were extracted from circular annuli extending to a radius of
$14\asec$ centred on the nucleus for obs. IDs 11513 to 14974.  The
radial range from these observations was limited by the subarray
selection and therefore spectra were also extracted from obs. ID 352
to extend the radial profiles to $\sim300\asec$.  Although there has been a significant decrease in the effective area at low energies between these observations due to the contaminant build up (\citealt{Marshall04}), the surface brightness profiles are consistent in the region of overlap (section A3).  Background spectra
were generated from the blank sky backgrounds for each source spectrum
and appropriate responses and ancillary responses were produced.  The
spectra were restricted to the energy range $0.5-7\keV$ and grouped
with a minimum of 25 counts per energy bin.  For obs. IDs 11513 to
14974, the source was located at a similar position on the same chip
and therefore the responses were comparable.  We summed together the
spectra and background spectra for each annulus and averaged the
response files with a weighting determined by the fraction of the
total counts in each observation.  The summed spectra contained a
minimum of $3000$ counts in each of the narrowest $2\asec$ radial bins
and over $40,000$ in the broader radial bins.

The summed spectra covering radii out to $14\asec$ and the obs. ID 352
spectra from 14 to $300\asec$ radius were deprojected with the
model-independent spectral deprojection routine \textsc{dsdeproj}
(\citealt{SandersFabian07}; \citealt{Russell08}).  Assuming spherical
symmetry, \textsc{dsdeproj} takes the background-subtracted spectra
and performs a geometric deprojection starting from the outermost
annulus and subtracting the projected emission off each successive
annulus (\citealt{Fabian81}; \citealt{Kriss83}).  The routine was
modified to include a correction for the regions excluded due to point
sources and a correction for the difference in effective area between
obs. ID 352 and the summed spectra.

The resulting deprojected spectra were each fitted in \textsc{xspec}
with absorbed single and two-temperature \textsc{apec} models
(\citealt{Balucinska92}; \citealt{Smith01}).  The cluster redshift was
fixed to $z=0.0044$ and the absorption of the cluster gas
  component was fixed to the Galactic column density
(\citealt{Kalberla05}), consistent with previous analyses
(eg. \citealt{FormanM8707}; \citealt{Million10}; \citealt{Werner10}).
The temperature, metallicity and normalization parameters were left
free and the $\chi^2$-statistic was minimised for the spectral fit.
For the two-temperature spectral fits, the metallicity was tied
between the two components.  An additional absorbed power-law
component was included to account for the nuclear PSF contribution in
the spectral fit to the summed spectra.  If the power-law parameters
are left free, the contribution of the nuclear PSF is incorrectly
interpreted in the spectral fitting as hot cluster emission.  The
parameters of the power-law component were therefore fixed to the
best-fit values determined from spectral fitting to deprojected
ChaRT/\textsc{marx} simulated spectra (eg. \citealt{Russell10};
\citealt{Siemiginowska10}).  We also generated a deprojected electron
density profile with finer radial binning from the PSF and background
subtracted surface brightness profile incorporating the temperature
and metallicity variations (eg. \citealt{Cavagnolo09}).  Finally, the
use of the summed spectra was verified by comparing the spectral fit
to the summed projected spectra with a simultaneous fit to all of the
individual projected spectra in \textsc{xspec}.  The best-fit
parameters for each case were found to agree to within the errors.


\section{Results}
\label{sec:results}

\begin{figure}
\centering
\includegraphics[width=0.98\columnwidth]{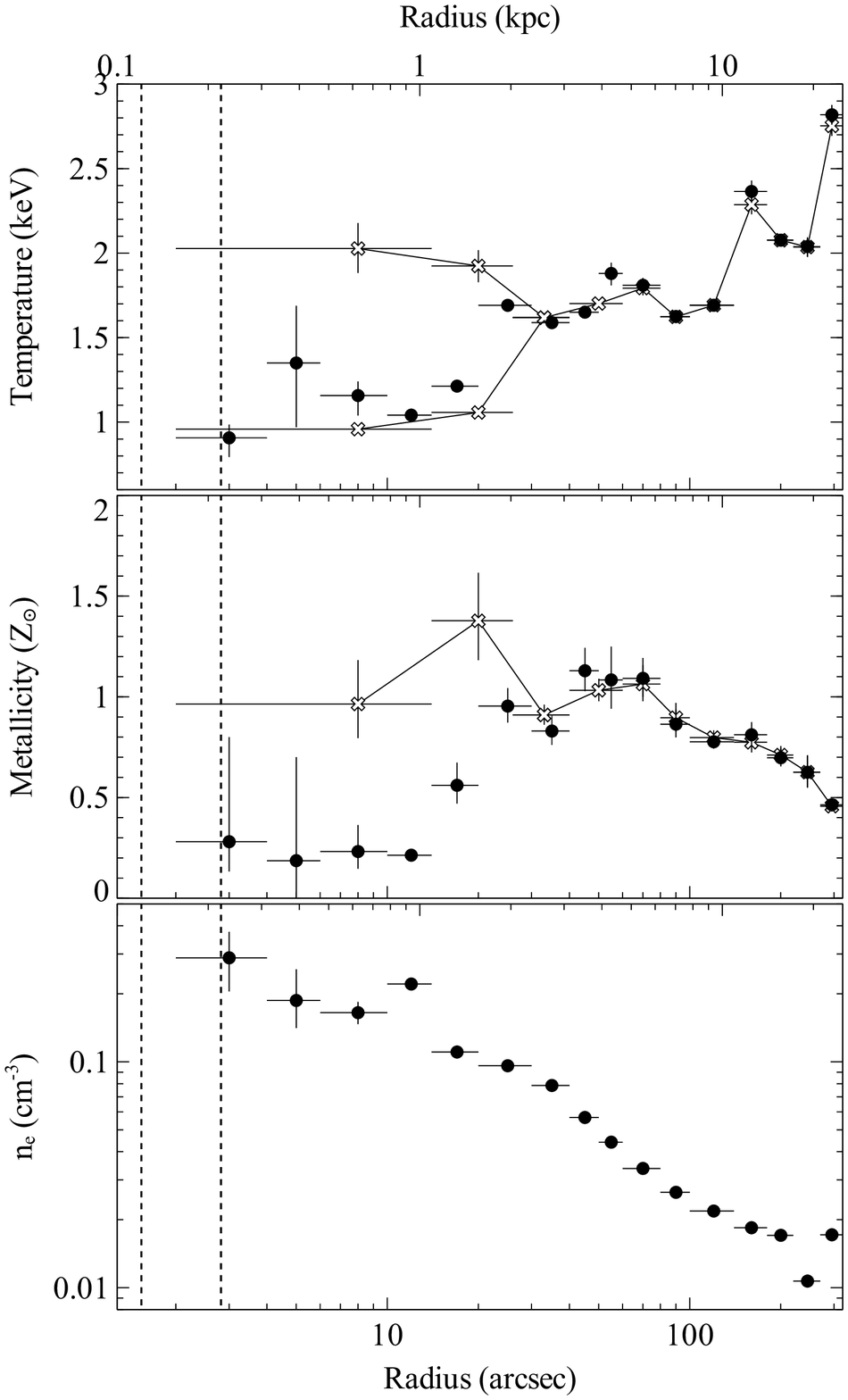}
\caption{Deprojected temperature, metallicity and density profiles comparing a single temperature spectral model (solid circles) and a two temperature spectral model (open crosses).  The radial range for the Bondi radius is shown by the vertical dashed lines.}
\label{fig:deproj1vs2temp}
\end{figure}

\begin{figure}
\centering
\includegraphics[width=0.98\columnwidth]{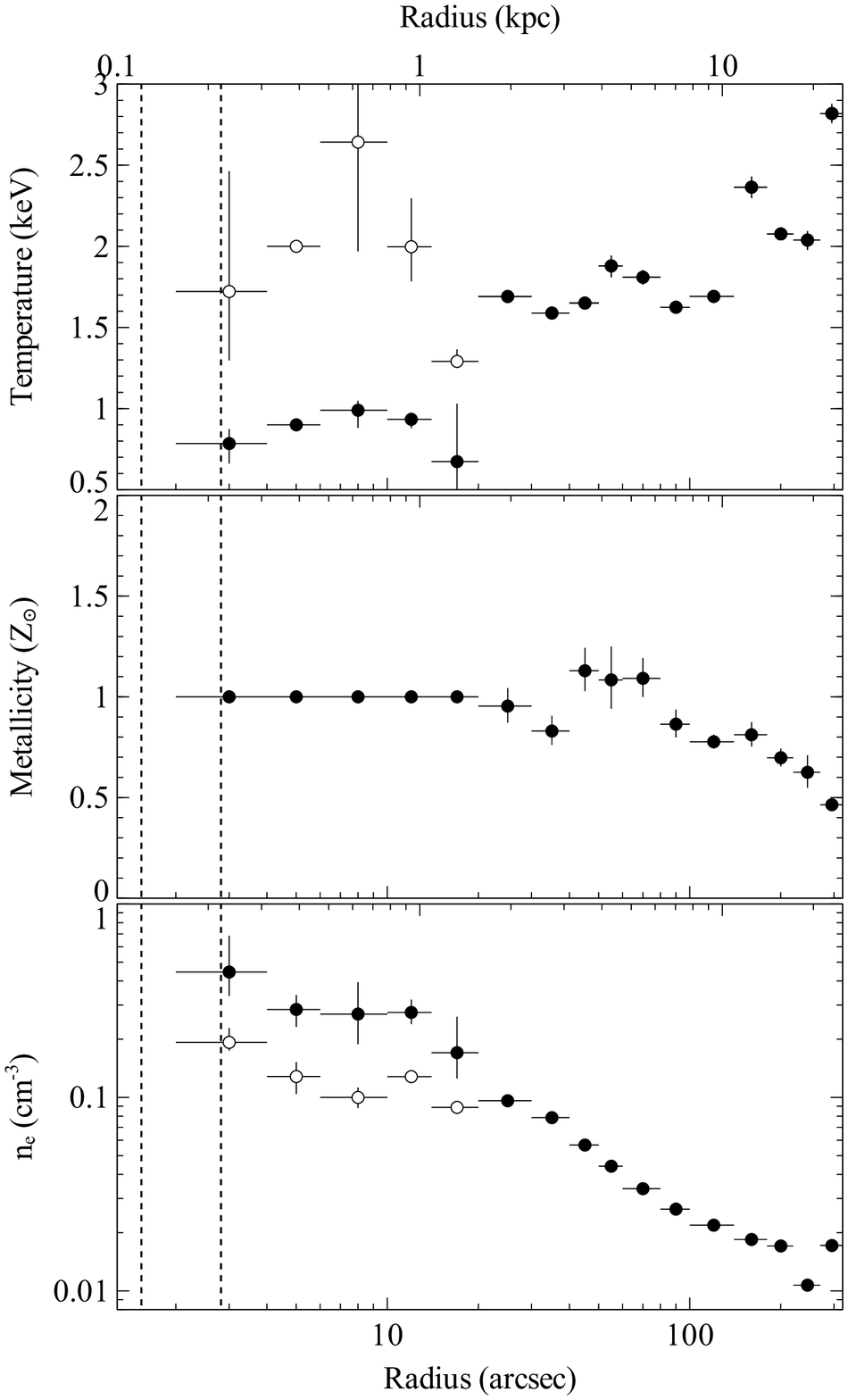}
\caption{Deprojected temperature, metallicity and density profiles for a two temperature spectral model.  The finer radial binning was achieved by fixing the metallicity inside a radius of $1.5\kpc$.  The density was calculated by assuming pressure equilibrium between the two temperature components.  The radial range for the Bondi radius is shown by the vertical dashed lines.}
\label{fig:deprojfine}
\end{figure}

Using a ChaRT/\textsc{marx} simulation to subtract the nuclear PSF
emission, we are able to determine the gas density down to $1\asec$
radius from the nucleus and the temperature down to $2\asec$ radius.
Fig. \ref{fig:deproj1vs2temp} shows the deprojected single temperature
best-fit results using radial bins of $2-4\asec$ width within the
central $14\asec$ ($1\kpc$) region.  The deprojected temperature drops
from $1.88^{+0.06}_{-0.07}\keV$ at $5\kpc$ ($64\asec$) to
$0.91^{+0.08}_{-0.11}\keV$ at $0.2\kpc$ ($2.5\asec$).  The coolest
X-ray gas is concentrated in the cluster centre
(eg. \citealt{Werner10}) producing a particularly steep drop in
temperature at $1.6\kpc$ ($20\asec$).  The bright, cool rim of the SE
cavity is located at a radius of $1\kpc$ and coincides with an
increase in the gas density by a factor of 2.  The density drops at a
radius of $\sim0.5\kpc$ due to the SE cavity and this produces larger
errors on the deprojected profiles due to the reduced number of counts
in this region.  The apparent increase in the deprojected temperature
at $0.3-0.8\kpc$ radius is therefore not significant and the profile
is consistent with a modest decline in temperature towards the cluster
centre.  However, it is also clear that a single temperature fit is
not adequate and multi-temperature gas is present requiring a
two-temperature spectral model.  The metal abundance peaks at
$1.1\pm0.1\Zsun$ at a radius of $4\kpc$ and then appears to drop
rapidly in the cluster centre.  This drop can be caused by the Fe bias
that occurs when a single temperature model is fitted to a
multi-temperature medium (eg. \citealt{Buote00}).

\subsection{Two temperature models}

Fig. \ref{fig:deproj1vs2temp} shows the deprojected temperature and
metallicity profiles produced by fitting two-temperature models to the
spectra extracted in broader radial bins.  Within $2\kpc$, a second
temperature component provides a better representation of the data with an F-test probability $<0.01\%$ that the second component is detected by chance.  The best-fit
temperature values are $1.98\pm0.09\keV$ and $1.00\pm0.02\keV$.  The
metallicity now appears to be consistent with $1\Zsun$ in the cluster
centre.  By fixing the metallicity to $1\Zsun$, a two-temperature
model was also fitted to the spectra extracted from the cluster centre, within $2\kpc$, in narrower radial bins.  Fig. \ref{fig:deprojfine} shows the
resulting deprojected profiles where multi-temperature gas was detected
significantly within $2\kpc$.  In addition to the metallicity
constraint, the temperature parameters were fixed to $0.9\keV$ and
$2\keV$ in the $0.3-0.45\kpc$ radial bin where the cavity increased
the uncertainties.  Only the \textsc{apec} normalization parameters
were left free for this region.  The gas temperatures for the two
components appear roughly constant within $1\kpc$, although
there could be a modest decrease in the lower temperature component.
No evidence was found for a temperature increase at small radii due to
the gravitational influence of the black hole, although this may occur at smaller radii within $0.15\kpc$.

The temperature profile down to a radius of $2\asec$ was not
significantly affected by the estimated uncertainty in the PSF
subtraction.  In section \ref{sec:dataanalysis}, the PSF-subtracted
cluster surface brightness profiles from obs. ID 1808 and the summed
obs. IDs 11513 to 14974 were compared and found to agree within the
errors.  However, the cluster surface brightness in the $1-2\asec$
radial bin is also consistent with a $\sim15\%$ difference between the
two epochs, which may be due to minor inaccuracies in the PSF
subtraction such as an underestimation of the pileup
  fraction.  The nuclear PSF flux is significantly lower in obs. ID
1808 therefore it is more likely that the PSF contribution has been
underestimated in the summed observations.  To be conservative, we
have therefore estimated the impact on the measured gas properties if
the PSF flux is higher by $20\%$.  This will increase the
  oversubtraction within $1\asec$ where the flux has been reduced by
  pileup but could improve the subtraction of the PSF wings.  No
significant change in the temperature profile was found for this
increase in the PSF flux.  We also note that no significant variation
in the temperature was found if the PSF flux was instead reduced by
$20\%$.  Within $2\asec$, the nuclear PSF dominates over the cluster
emission and the uncertainty in the faint hard X-ray emission above
$3\keV$ produced a large uncertainty in the gas temperature.  The gas
temperature profile is therefore restricted to radii greater than
$2\asec$.


\subsection{Density profile}

The coolest component of the multi-temperature gas in M87 likely
consists of small blobs and filaments whilst the $2\keV$ component
fills the volume, excluding the cavities (\citealt{Young02};
\citealt{Sparks04}; \citealt{Werner10}).  By assuming that the
different gas phases are in pressure equilibrium, the filling factor
of the cooler gas component can be calculated and used to determine
the gas density.  Note that we have ignored magnetic pressure, which is likely required to support extended filaments of cold gas (\citealt{Fabian08}).  The gas density of both temperature components
continues to rise towards the cluster centre and the profile appears
to steepen inside $0.5\kpc$ radius.  A similar increase is seen in
Fig. \ref{fig:deproj1vs2temp} for the single temperature density
profile.  To determine if this steepening is significant, we have also
produced a deprojected density profile with narrower $1\asec$ radial
bins from the PSF and background-subtracted surface brightness profile
(see section \ref{sec:clusterspec}).

\begin{figure}
\centering
\includegraphics[width=0.98\columnwidth]{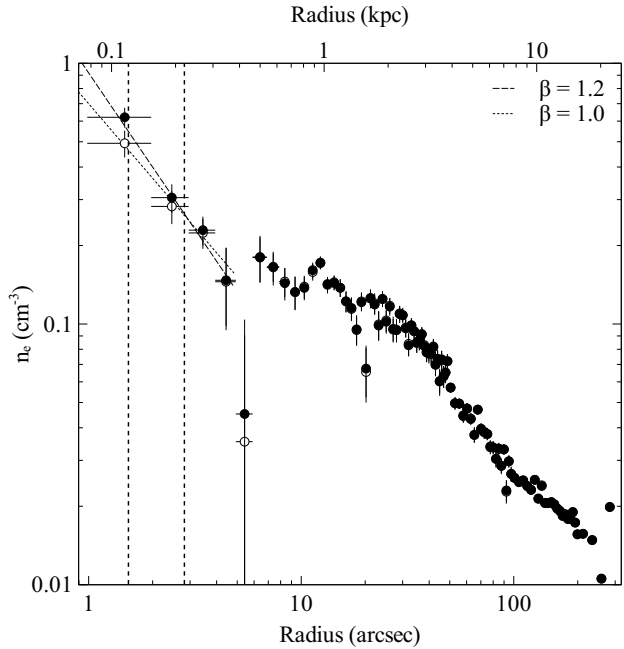}
\caption{Deprojected density profile.  The nuclear flux contribution has been subtracted off using a ChaRT simulation (solid circles).  The open circles show the deprojected density profile if the nuclear flux was increased by $20\%$.  The radial range for the Bondi radius is shown by the vertical dashed lines.}
\label{fig:deprojdensity}
\end{figure}

Fig. \ref{fig:deprojdensity} shows the higher spatial resolution
density profile which extends in to $1\asec$.  The density profile
steepens within $0.3\kpc$ and peaks at a density of
$0.62\pm0.05\pcmcu$.  Measurement of the gradient is however
complicated by the cavity at $\sim0.5\kpc$ radius, where the emitting
volume is overestimated and therefore the density is underestimated.  Due to the cavity, the majority of the emission at this radius is projected.  Therefore the density value is also sensitive to the PSF subtraction despite this accounting for only $\sim2\%$ of the total emission.  The nuclear PSF accounts for roughly half of the total emission
in the $1-2\asec$ radial bin and therefore any inaccuracy in the
subtraction could have a significant effect here
(Fig. \ref{fig:psfprepostsub}).  Fig. \ref{fig:deprojdensity} also
shows the resulting density profile if the nuclear PSF flux is
increased by $20\%$.  The peak density drops to $0.49\pm0.06\pcmcu$
and the gradient of the profile inside $0.3\kpc$ radius is shallower.
Fig. \ref{fig:deprojdensity} shows two possible lines of best-fit to
the density profile at the centre of M87.  The best-fit power-law
slope of $-1.2\pm0.2$ assumed that the PSF subtraction is sufficiently
accurate and excluded all points beyond a radius of $0.3\kpc$.  The
shallower best-fit slope of $-1.0\pm0.2$ was determined using the
increased PSF subtraction.  The density profile is therefore
consistent with $\rho{\propto}r^{-1}$ within the Bondi radius.

\subsection{Cooling time and dynamical time profiles}

\begin{figure}
\centering
\includegraphics[width=0.98\columnwidth]{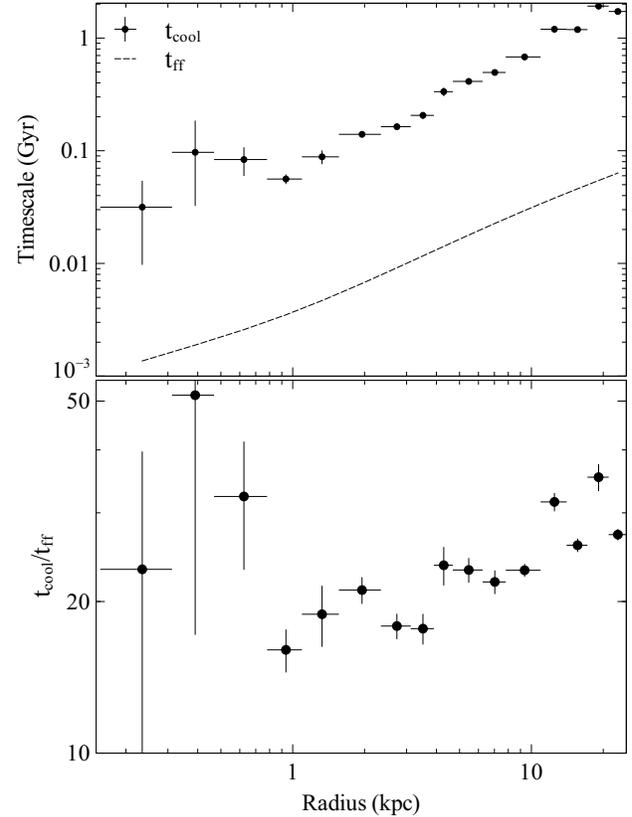}
\caption{Upper panel: the radiative cooling time profile for a single temperature component model and the dynamical free-fall time profile.  Lower panel: the ratio of cooling time to free-fall time.}
\label{fig:tcoolovertff}
\end{figure}

Multi-temperature structure is expected to form in the ICM when the
radiative cooling time of the gas becomes short with respect to the
gas dynamical time.  Within $2\kpc$, a second temperature component is
significantly detected in the hot gas and this is also coincident with
the extent of the brightest H$\alpha$ filaments (\citealt{Young02};
\citealt{Sparks04}).  We have therefore calculated the radiative
cooling time and dynamical time profiles to determine if this observed
structure is consistent with the development of local thermal
instabilities.  The radiative cooling time, $t_{\mathrm{cool}}=\left(3/2\right)nkT/n^2\Lambda$,
was determined from the single component temperature and density
profiles.  For the two component spectral model, the cooling time is
approximately a factor of two longer for the hotter component and a
factor of $2-3$ shorter for the colder component.  The dynamical time
was calculated using the mass profile for M87 of \citet{Romanowsky01}
(see also \citealt{Churazov08}).  Fig. \ref{fig:tcoolovertff} shows
the cooling time, the free-fall time and the
$t_{\mathrm{cool}}/t_{\mathrm{ff}}$ profiles for M87.  The radiative cooling time drops to $3\pm2\times10^7\yr$ at $0.2\kpc$ and is even shorter for the cool X-ray temperature component at $\left(1.1\pm0.8\right)\times10^7\yr$.  The
ratio reaches a minimum of $t_{\mathrm{cool}}/t_{\mathrm{ff}}=16\pm2$
at a radius of $0.9\kpc$ and is comparably low to a radius of
$\sim4\kpc$.  

\subsection{Bondi radius and accretion rate}
\label{sec:Bondi}

The SMBH should accrete from the surrounding hot
atmosphere at a rate determined by its mass and the gas
density and temperature at the radius where the SMBH's gravitational
influence dominates (\citealt{Bondi52}).  Under the assumptions of spherical symmetry and
negligible angular momentum, the Bondi accretion rate is given by

\begin{equation}
\frac{\dot{M}_{\mathrm{B}}}{\Msunpyr}=0.012\left(\frac{k_{\mathrm{B}}T}{\mathrm{keV}}\right)^{-3/2}\left(\frac{n_e}{\pcmcu}\right)\left(\frac{M_{\mathrm{BH}}}{10^9\Msun}\right)^2,
\end{equation}

\noindent where $T$ the gas temperature, $n_e$ is the gas electron
density and $M_{\mathrm{BH}}$ is the SMBH mass.  An adiabatic index
$\gamma=5/3$ is assumed.  The Bondi radius can also be expressed as

\begin{equation}
\frac{r_{\mathrm{B}}}{\mathrm{kpc}}=0.031\left(\frac{k_{\mathrm{B}}T}{\mathrm{keV}}\right)^{-1}\left(\frac{M_{\mathrm{BH}}}{10^{9}\Msun}\right).
\end{equation}

\noindent Recent gas-dynamical and stellar-dynamical black hole mass estimates for M87 differ by a factor
of two from $M_{\mathrm{BH}}=6.6\pm0.4\times10^{9}\Msun$
(\citealt{Gebhardt11}) to
$M_{\mathrm{BH}}=3.5^{+0.9}_{-0.7}\times10^{9}\Msun$
(\citealt{Walsh13}).  Therefore, our estimates of the Bondi radius and
accretion rate are expressed as a range covering these two
values of the black hole mass.  

For a single temperature spectral model, the temperature falls to
$0.91^{+0.08}_{-0.11}\keV$ in the centre of M87 and therefore the
Bondi radius $r_{\mathrm{B}}=0.12-0.22\kpc$ ($1.5-2.8\asec$).  The
electron density at the Bondi radius ranges between $0.31\pm0.04\pcmcu$
and $0.62\pm0.05\pcmcu$, which gives Bondi accretion rates
$\dot{M}_{\mathrm{B}}=0.1-0.2\Msunpyr$.  Assuming a standard
$\sim10\%$ efficiency and accretion onto the SMBH at the Bondi rate,
the Bondi accretion power is $P_{\mathrm{B}}=0.5-1\times10^{45}\ergps$.  Bondi accretion
therefore would be sufficient to fuel the mechanical power output from
the jet of $P_{\mathrm{jet}}=8^{+7}_{-3}\times10^{42}\ergps$ (eg. \citealt{Bicknell99}; \citealt{DiMatteo03};
\citealt{Rafferty06}; \citealt{Russell13}).  If we consider only the
higher, volume-filling temperature component of the two temperature spectral
model, the Bondi accretion rate would be a factor of $\sim5$ lower but
still sufficient to power the jet.  For a nuclear bolometric
luminosity $L_{\mathrm{bol}}\sim2\times10^{41}\ergps$, the Bondi
accretion power is also far greater than the nuclear radiation losses
(\citealt{Reynolds96}; \citealt{DiMatteo03}).  Both the accretion flow
and the jets are therefore radiatively inefficient
(\citealt{DiMatteo03}), although the accretion rate may be much lower at the event horizon.

\section{Discussion}
\label{sec:discussion}

The gas density and temperature profiles within the accretion radius
may in principle reflect the gravitational influence of the central
SMBH and permit us to distinguish between different accretion flow models.  M87 is
one of only a few systems with a Bondi radius that can be resolved by
\textit{Chandra} but the analysis is complicated by the bright central
point source and jet emission.  By stacking short frame-time
observations of M87 taken after the HST-1 flaring period and using a
ChaRT/\textsc{marx} simulation to subtract the PSF, we determined the
gas density to within the Bondi radius and traced the multi-temperature
structure on these spatial scales.  


\subsection{Hot gas accretion}

Within $2\kpc$, two temperature components are significantly detected,
which are consistent with constant values of $2\keV$ and $0.9\keV$
down to a radius of $0.15\kpc$.  No evidence was found for an increase
in gas temperature within $\sim0.25\kpc$ radius where the SMBH is
expected to gravitationally heat the ambient gas, although
  this may occur at small radii within $0.15\kpc$
(eg. \citealt{Brighenti99}; \citealt{Quataert00}).  Evidence for
central temperature increases on this scale has been found in several
nearby elliptical galaxies (\citealt{Pellegrini03Sombrero};
\citealt{Machacek06Nuc}; \citealt{Humphrey08}; \citealt{Pellegrini12})
and, in NGC3115, the hotter, volume-filling component of a two
temperature model appears to increase to $\sim1\keV$ towards the
centre (\citealt{Wong14}).

However, the hot atmosphere at the centre of M87 has been
significantly disturbed by the AGN.  Several X-ray cavities are
coincident with radio jet emission in the core of M87, including one
at a radius of only $0.5\kpc$ to the SE of the nucleus in the
counter-jet direction (\citealt{Young02};
\citealt{FormanM8705,FormanM8707}).  This activity will generate
shocks, stir up the gas and cause departures from hydrostatic
equilibrium (eg. \citealt{FormanM8705}; \citealt{Churazov08}).  The
counter-jet cavity corresponds to a region of apparent low pressure
and high projected temperature (\citealt{Million10};
\citealt{Werner10}).  The hotter $\sim2\keV$ component of the
multi-temperature ICM inside a radius of $\sim2\kpc$ may correspond to
gas heated by this recent jet outburst.  The multi-temperature
  structure could therefore be due to hotter, shock-heated regions and
  dense clumps of cooling gas but this cannot be distinguished from a
  more homogeneous multi-temperature distribution by the
  azimuthally-averaged temperature profiles.  Finer spatial binning
of the temperature profile extending in to $1\asec$ radius, excluding
the region affected by the AGN activity and tracing any
  azimuthal variation may be possible with additional observations if
the nuclear flux continues to decline.  However, with the existing
data, the central temperature gradient is likely affected by the AGN
activity and may not reflect the underlying gravitational potential.


The density profile was extracted using finer spatial bins, which
extend in to $1\asec$, and is also more robust to uncertainties in the
PSF subtraction.  The profile flattens within $\sim3\kpc$ with a ridge
of increased density at $1\kpc$ ($\sim12.5\asec$) corresponding to the
bright rim of the inner bubble.  From the density and temperature
profiles, we calculated the Bondi accretion rate
$\dot{M}_{\mathrm{B}}=0.1-0.2\Msunpyr$ across the Bondi radius located
at $0.12-0.22\kpc$ ($1.5-2.8\asec$), depending on the black hole mass.
For accretion onto the SMBH at the Bondi rate and assuming
10\% efficiency, the accretion rate is roughly two orders of magnitude
greater than that required by the mechanical power output of the jet.
The density profile for a Bondi flow is given by
$\rho{\propto}r^{-\beta}$, where $\beta=3/2$ for
$r{\ll}r_{\mathrm{B}}$.  Close to the Bondi radius, on the scales
probed by the \textit{Chandra} observations, a Bondi flow has a
shallower gradient, $\beta<1$ (\citealt{Bondi52}; \citealt{Narayan94};
\citealt{Quataert00}).  Excluding the region affected by the cavity,
the density profile inside the Bondi radius in M87 is consistent with
$\rho{\propto}r^{-1}$, which may indicate the onset of a Bondi flow.
However, the ADIOS and CDAF models are also supported by a shallow
density profile and cannot be ruled out by the existing observations.
Alternatively, the density profile could reflect the radio jet
activity rather than the underlying gravitational potential.
\textit{Chandra} observations of NGC\,4889, one of two BCGs in the
Coma cluster with a black hole mass of $2\times10^{10}\Msun$
(\citealt{McConnell11}), show a central cavity inside $0.6\kpc$ radius, which is comparable to the recent outburst in M87
(\citealt{Sanders14}).  The density distribution is consistent with no
X-ray emission within this radius suggesting that much of the material
has been driven out by the jet.  Although we cannot
rule out the different models, it is not clear that the hot gas
structure across the Bondi radius is dictated by the gravitational
potential of the SMBH and the classical Bondi accretion rate may not
reflect the accretion rate onto the SMBH.

\subsection{Hot outflow?}


Faraday rotation measurements can provide constraints on the accretion
rate much closer to the SMBH on scales of tens of Schwarzschild radii
($r_{\mathrm{s}}$).  For M87, \citet{Kuo14} determined an upper limit
on the accretion rate at $21r_{\mathrm{s}}$ of
$9.2\times10^{-4}\Msunpyr$.  This limit is at least two orders of
magnitude below the Bondi rate and it is consistent with the density
profile $\rho{\propto}r^{-1}$, determined by \textit{Chandra}, extending
in to this radius.  However, this analysis assumes that the observed
Faraday rotation originates in the accretion flow, the density profile
follows a power-law with $\beta\leq3/2$ and an equipartition strength
magnetic field.

Observations of Sgr A$^*$ and NGC3115 are consistent with the onset of
an accretion flow inside the Bondi radius and a reduced accretion rate
onto the SMBH.  Using a $1\Ms$ \textit{Chandra} observation of
NGC3115, \citet{Wong14} found a shallow density profile,
$\rho{\propto}r^{-1}$, and a hot gas component with increasing
temperature within the Bondi radius reflecting the gravitational
potential of the SMBH.  For Sgr A$^*$, \citet{Wang13} showed that the
density profile is consistent with $\rho{\propto}r^{-1/2}$, assuming a
temperature profile $T{\propto}r^{-1}$.  Faraday rotation measurements
of Sgr A$^*$ provide an upper limit on the accretion rate at
$r{\leq}100r_{\mathrm{s}}$ (\citealt{Bower03};
\citealt{Marrone06,Marrone07}; \citealt{Macquart06}).  When combined
with the X-ray measurements at the Bondi radius, this suggests that
the flat density profile covers a broad radial range down to
$r\sim100r_{\mathrm{s}}$ indicating little material captured at the
Bondi radius reaches the SMBH (\citealt{Yuan03}; \citealt{Wang13}).

These observations are consistent with numerical simulations and
analytical models of hot accretion flows, with accretion rate
decreasing with decreasing radius so that the majority of the material
is lost between the Bondi radius and the event horizon
(eg. \citealt{Igumenshchev99}; \citealt{Stone99}; \citealt{Stone01};
\citealt{Hawley02}; \citealt{Yuan10}; \citealt{Begelman12}).  The
radial density profile becomes correspondingly flatter with
$\rho{\propto}^{-(0.5-1)}$.  Recent simulations by \citet{Yuan12}
covering a greatly expanded radial range found density profiles
described by $\rho{\propto}^{-(0.65-0.85)}$ for $r>10r_{\mathrm{s}}$,
favouring the ADIOS model where the majority of the mass is blown out
in a wind (\citealt{Yuan12}).  

If the accretion rate onto the SMBH in M87 is similarly reduced from
the Bondi rate to $\dot{M}_{\mathrm{BH}}=9.2\times10^{-4}\Msunpyr$ (at
$21r_{\mathrm{s}}$, \citealt{Kuo14}), the accretion power is
$P_{\mathrm{BH}}=5\times10^{42}\ergps$, where an efficiency of $10\%$
has been assumed.  This is two orders of magnitude below the Bondi
accretion power and close to the mechanical power output generated by
the jet $P_{\mathrm{jet}}=8^{+7}_{-3}\times10^{42}\ergps$.  The bulk
of the mass flow will presumably be lost in a wind, which will deposit
mass and momentum in the surrounding ambient gas
(eg. \citealt{Proga03}; \citealt{Ostriker10}).  This will reduce the
inflow, controlling the growth of the SMBH and making the accretion
flow time variable.  Calculation of the Bondi rate also assumes that
the hot gas at the Bondi radius is falling toward the SMBH.  Instead,
the temperature profile indicates the hot gas dynamics may not be
dictated by the SMBH and the metallicity structure and ionized gas
dynamics suggest outflow.  The alignment of cool, metal-rich gas with
the radio jet and cavity axes is consistent with a jet-driven outflow
of gas uplifted from the cluster centre (\citealt{Belsole01}; \citealt{Matsushita02}; \citealt{Molendi02}; \citealt{Simionescu08};
\citealt{Kirkpatrick11}).  The most recent AGN outburst generated the
inner radio bubbles $\sim10^7\yr$ ago and uplifted roughly
$5\times10^8\Msun$ of gas to $\sim20\kpc$ radius
(\citealt{Simionescu08}).

The state of the hot gas in M87 differs from the cold gas disk, which is clearly
dictated by the SMBH (\citealt{Harms94}; \citealt{Macchetto97};
\citealt{Walsh13}).  The central ionized gas disk rotates at
velocities from $-500$ to $+500\kmps$ with a mass of
$\sim4\pm1\times10^{3}\Msun$ inside $1\asec$ radius (\citealt{Ford94};
\citealt{Harms94}; \citealt{Macchetto97}).  However, optical and UV
observations of extended dusty gas filaments found broad emission
lines with both blue and redshifted non-Keplerian components at
several hundred km/s with respect to the systemic velocity
(\citealt{Tsvetanov98}; \citealt{Ford99}).  Several filaments appear
to connect to the disk and the kinematics of the inner filaments are
consistent with a bi-directional wind blowing from the disk
(\citealt{Sparks93}; \citealt{Ford99}).  The preponderance of evidence
supports an outflow from the hot gas and the disk of cold gas.

\subsection{Cold inflow?}

Most of the circumnuclear gas in the rotating disk is presumably in
the form of molecular clouds but currently only tentative detections
and upper limits typically of several $\times10^{6}\Msun$ have been
found (\citealt{Braine93}; \citealt{Combes07}; \citealt{Tan08};
\citealt{Salome08}).  Loss of angular momentum through collisions
between clouds and the wind blowing off the disk could allow this gas
to flow inwards and fuel the SMBH.  In the `cold feedback' model, these
circumnuclear gas disks can be fed by overdense, thermally unstable
blobs of gas that cool rapidly from the surrounding hot atmosphere and
lose angular momentum through drag in the ICM to accrete onto the disk
(\citealt{PizzolatoSoker05}; \citealt{Nulsen86};
\citealt{Pizzolato10}).  Short central radiative cooling times in the
hot ICM are strongly correlated with the presence of extended cool gas
filaments suggesting that these originate in cooling of the hot gas
(\citealt{Heckman81}; \citealt{Cowie83}; \citealt{Peres98}).

Fig. \ref{fig:deproj1vs2temp} shows that the hot gas in M87 is
multi-temperature within $2\kpc$, which coincides with the extent of
the brightest H$\alpha$ filaments (\citealt{Young02};
\citealt{Sparks04}; \citealt{Werner10}).  Numerical simulations of
thermal instability in the ICM suggest that this multi-temperature structure
will form when the gas cooling time drops to a factor of ten greater
than the dynamical free-fall time,
$t_{\mathrm{cool}}/t_{\mathrm{ff}}{\leq}10$ (\citealt{McCourt12};
\citealt{Sharma12}; \citealt{Voit14}).  The ratio of the radiative
cooling time to the free-fall time in M87 drops to
$t_{\mathrm{cool}}/t_{\mathrm{ff}}=16\pm2$ at $0.9\kpc$ and is
comparably low to a radius of $\sim4\kpc$.  This appears inconsistent
with the criterion for multi-temperature gas where
$t_{\mathrm{cool}}/t_{\mathrm{ff}}{\leq}10$ but \citet{Voit14} suggest
that AGN feedback can cause the minimum ratio value to fluctuate and a
broader range of values may be acceptable.  \citet{PizzolatoSoker05}
show that, in addition to short cooling times and a suppressed rate of
thermal conduction, the ambient density profile must be shallow for
cold feedback to occur.  Overdense, rapidly cooling gas will be
prevented from reaching an equilibrium position in the ICM and can
cool to low temperatures.  The most significant flattening of the
density profile in M87 occurs at $\sim3\kpc$ radius, which also
encompasses the brightest H$\alpha$ emission and multi-temperature ICM
structure.

The radiative cooling time drops to only $10^7\yr$ for the cooler
X-ray temperature component inside $4\asec$ radius.  There are clear
associations between the soft X-ray emission and the optical line
filaments (\citealt{Sparks04}) and, whilst the innermost
filaments are outflowing, the outer filaments appear to be falling in
towards the disk (\citealt{Sparks04}).  These overdense, rapidly
cooling gas clouds can decouple from the ICM and nearly free-fall to
the cluster centre (\citealt{Pizzolato10}).  At $0.2\kpc$ radius, the
cool X-ray gas component represents $\sim20\%$ of the total X-ray gas mass
and could lose angular momentum to the hot gas component if this is
not rotating.  The cooler X-ray temperature component could therefore
provide a source of cold gas clouds within the Bondi radius with a mass
deposition rate of $\sim0.1\Msunpyr$.  This would produce sufficient
fuel to power the most recent radio jet outburst.  The X-ray mass
deposition rate measured by the \textit{XMM-Newton} RGS in a
$1.1\amin$ wide aperture ($5.2\kpc$) is $\dot{M}=0.9\Msunpyr$ for gas
cooling radiatively from $1\keV$ to $0.5\keV$ and
$\dot{M}<0.06\Msunpyr$ below $0.5\keV$ (\citealt{Werner10}).  However,
if the coolest X-ray gas mixes with the cold H$\alpha$-emitting
filaments, which may be promoted by the AGN activity, then there could
be significant non-radiative cooling of the gas (\citealt{Fabian02};
\citealt{Soker04}; \citealt{Sanders10}).  Taken together, the data are
consistent with cold feedback where gas cooling from the hot ICM feeds
the circumnuclear gas disk.

\section{Conclusions}
\label{sec:conclusions}

Using recent short frame-time \textit{Chandra} observations of M87
taken after the flaring period of the HST-1 jet knot, we have
subtracted the nuclear PSF to determine the temperature and density
structure of the hot gas within the Bondi accretion radius.  Within
$2\kpc$, we detect two significant temperature components, which are
consistent with constant values of $2\keV$ and $0.9\keV$ down to a
radius of $0.15\kpc$.  No evidence was found for the expected increase in the
gas temperature within a radius of $\sim0.25\kpc$, where the SMBH is
expected to gravitationally heat the surrounding gas.  This may occur
at smaller radii within $0.15\kpc$.  


From the temperature and density profiles, we calculated the Bondi
accretion rate of $0.1-0.2\Msunpyr$ across the Bondi radius at
$0.12-0.22\kpc$ ($1.5-2.8\asec$), depending on the black hole mass.
The density profile flattens inside $3\kpc$ and then steepens again
within $0.3\kpc$ ($4\asec$) to peak at \textbf{$0.5-0.6\pcmcu$}.
Inside the Bondi radius, the density profile is consistent with
$\rho{\propto}r^{-1}$.  Whilst a temperature rise within $0.15\kpc$
cannot be ruled out, the absence of the expected temperature increase
inside the Bondi radius indicates that the hot gas dynamics are not
dictated by the SMBH and, together with the shallow density gradient,
suggests that the classical Bondi rate may not reflect the SMBH's
accretion rate.  If this shallow density gradient extends inwards to
the SMBH, consistent with the Faraday rotation measurements for M87,
the accretion rate onto the SMBH could be two orders of magnitude
below the Bondi rate.  Whilst this is still sufficient to power the
recent radio jet outburst, assuming an efficiency of $10\%$, the
accretion flow will be significantly altered if a wind blows out the
majority of the mass.  The uplift of metal-rich gas along the radio
jet axis and broad optical and UV emission lines in the ionized gas
disk support the existence of an outflow in the hot gas and the cold
gas disk.

Clear associations exist between the coolest X-ray gas and the optical
line-emitting filaments at the centre of M87 and several outer
filaments appear to be infalling towards the disk.  Observations show
that the ICM has multiple temperature components inside a radius of
$2\kpc$, the radiative cooling time of the coolest X-ray gas drops to
$10^7\yr$ here and this rapidly cooling gas is coincident with the
extent of the bright H$\alpha$ emission.  The density profile also flattens
significantly at this radius, which is consistent with cold feedback
models that suggest a shallow density profile will prevent overdense,
cooling blobs from reaching an equilibrium position.  The cooler X-ray
temperature component could provide a source of cold gas clouds within
the Bondi radius and sufficient fuel to power the most recent radio
jet outburst.


\section*{Acknowledgements}

We thank the reviewer for helpful and constructive comments.  HRR and
ACF acknowledge support from ERC Advanced Grant Feedback.  BRM
acknowledges support from the Natural Sciences and Engineering Council
of Canada and the Canadian Space Agency Space Science Enhancement
Program.  AEB receives financial support from the Perimeter Institute
for Theoretical Physics and the Natural Sciences and Engineering
Council of Canada through a Discovery Grant.  Research at Perimeter
Institute is supported by the Government of Canada through Industry
Canada and by the Province of Ontario through the Ministry of Research
and Innovation.  We thank Alastair Edge for helpful discussions.  The
scientific results reported in this article are based on data obtained
from the Chandra Data Archive.

\section*{Appendix: Simulating the nuclear PSF}

\subsection*{A1 Pileup}


Two or more photons landing on the same ACIS detector region within
the same frame integration time will be detected as a single event.
Known as pileup (see eg. \citealt{Davis01}), the photon energies sum
to make a detected event of higher energy.  The source spectrum
hardens and the altered shape of the charge cloud distribution can
cause grade migration, where the event is misidentified as a cosmic
ray and excluded from analysis.  Most of the intensity of piled up
sources should still be recovered.  However, when the jet knot HST-1
was near its peak, strong pileup produced events that exceeded the
energy filter cutoff and were not telemetered to the ground resulting
in lost flux (\citealt{Harris09}).  The proximity of HST-1 to the
bright nucleus also produced the `eat-thy-neighbour' effect where
photons arriving within the same frame and $3\times3$ pixel grid are
considered part of the same event.  The photon with the most energy
then determines the recorded location for both (\citealt{Harris09}).
It was therefore important to determine the extent of any pileup in
the observations used and whether it was possible to accurately
subtract the nuclear PSF from the surrounding extended emission.

The pileup fraction was estimated using the ratio of `good' event
grades to `bad' grades in a region of $1\asec$ radius centred on the
emission peak.  Most of the nuclear emission comes from the
$0.5-2\keV$ energy band which, if piled up, increases the number of higher
energy events above $2\keV$ with `bad' grades.  We therefore used the
ratio of `bad' to `good' grades in the $2-5\keV$ energy band as an
estimate of the pileup fraction but note that this measure is
dependent on the nuclear spectrum and affected by onboard rejection.
This analysis was confirmed in section A3 with a \textsc{marx}
simulation of the nuclear PSF that included a statistical treatment of
pileup (\citealt{Davis01}).


Apart from obs. ID 352, all of the observations used have a short
frame integration time of $0.4\s$ to reduce pileup.  We estimate that
obs. ID 1808 has the lowest pileup fraction at $6\%$ and the more
recent observations taken since 2010 have a slightly higher pileup
fraction of $10-15\%$ because the nucleus was a factor of $\sim1.5-2$
brighter.  Obs. ID 352 has a standard frame integration time of
$3.2\s$ and is therefore strongly piled up by $\sim80\%$.  Following
\citet{Harris06}, the brightness of HST-1 was determined using a
detector-based measure of intensity ($\keVps$), which sums the photon
energies from the event 1 file with no grade filtering.  The HST-1
brightness drops from $0.3\keVps$ in 2010 to $0.2\keVps$ at the end of
2012.  These measurements are all more than an order of magnitude
below the $4\keVps$ limit above which \citet{Harris09} suggest that
more complex pileup problems occur.  The measurements also confirm
that HST-1 has returned to the 2000 brightness level before the major
flare and pileup is now only mild for observations with a frame
integration time of $0.4\s$.

\begin{figure*}
\begin{minipage}{\textwidth}
\centering
\includegraphics[width=0.45\columnwidth]{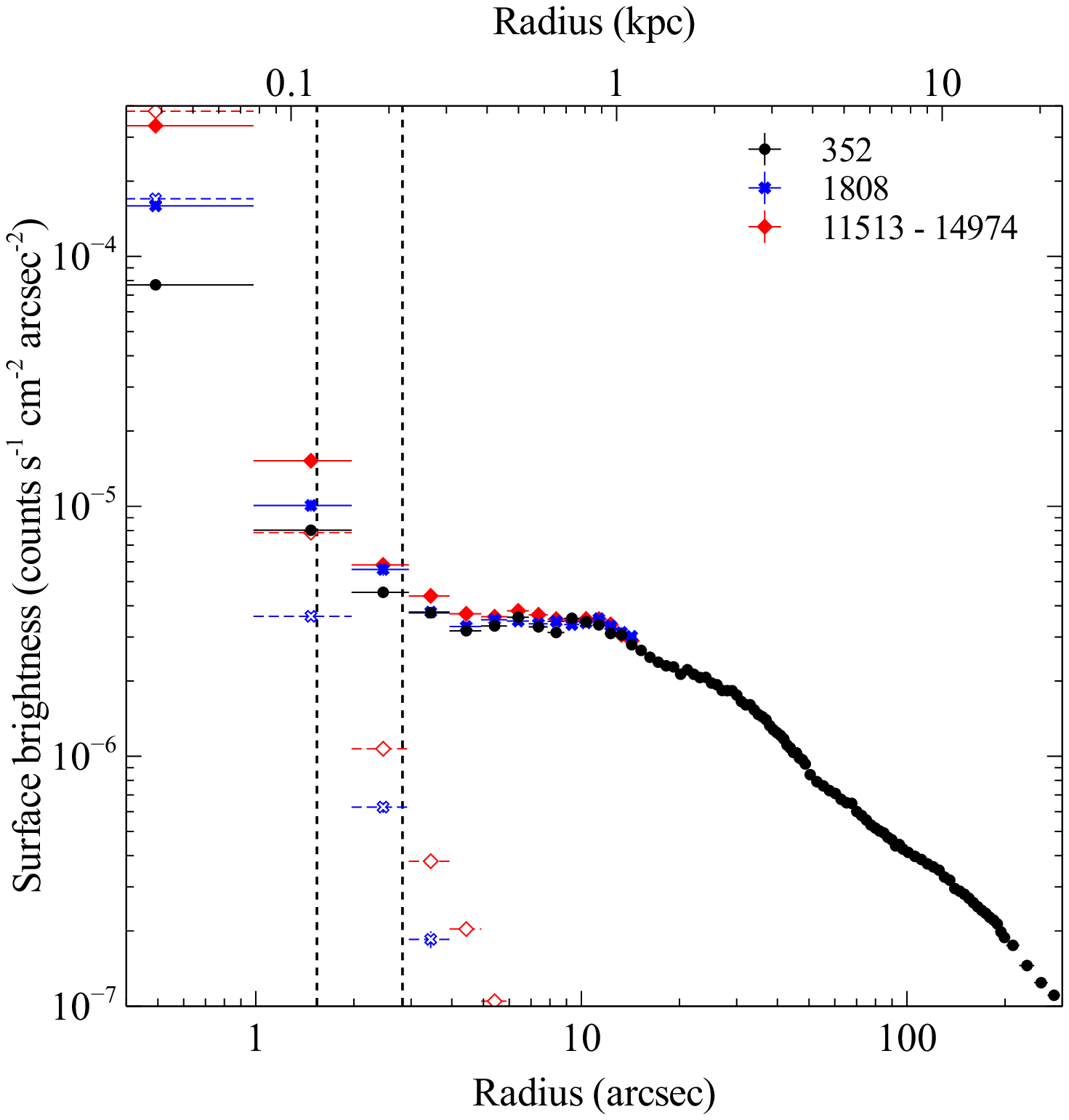}
\includegraphics[width=0.45\columnwidth]{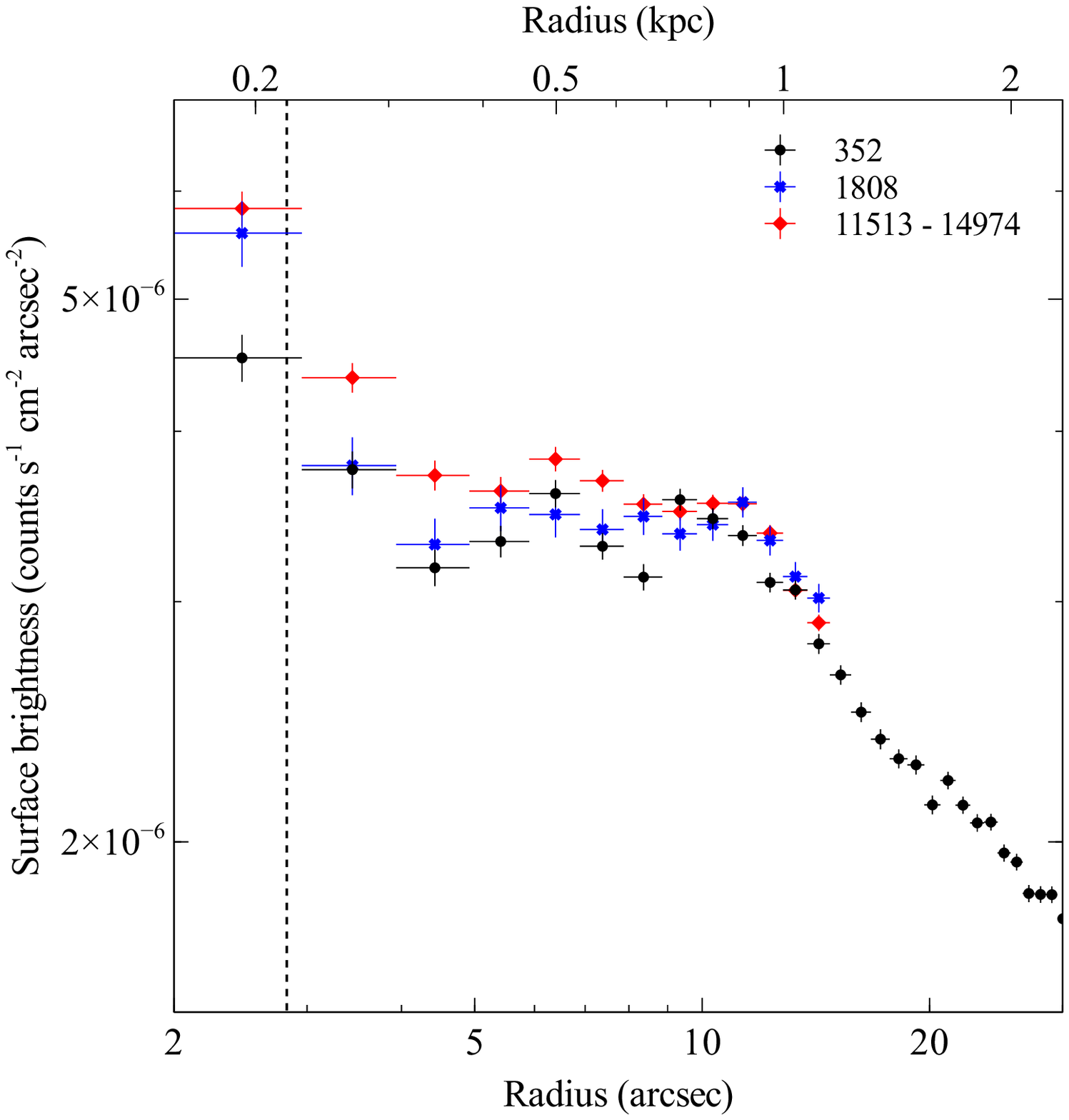}
\caption{Left: Background-subtracted surface brightness profiles in the energy band $0.5-7.0\keV$ for obs. ID 352, obs. ID 1808 and summed obs. IDs 11513 to 14974.  The open symbols show the predicted contribution to the total emission by the nuclear PSF from the ChaRT simulations for obs. IDs 11513 to 14974 (red) and obs. ID 1808 (blue).  The radial range for the Bondi radius is shown by the vertical dashed lines.  Right: Zoom in showing the region of overlap between the subarray and full frame datasets.}
\label{fig:fullradrange}
\end{minipage}
\end{figure*}

\subsection*{A2 Nuclear spectrum}
\label{sec:nucspec}


For each observation, we extracted a nuclear spectrum using a region
of $1\asec$ radius centred on the nuclear emission peak.  We
  assume that all emission within $1\asec$ is from the AGN.  However
  the gas emission is likely to be $\sim5\%$ of the total, depending
  on the slope of the gas density profile.  A corresponding
background spectrum, including the cluster emission, was extracted
using a surrounding annulus from $2-4\asec$.  The spectra were
restricted to the $0.5-7\keV$ energy band and appropriate responses
and ancillary responses were generated.  The spectra were fitted
individually in \textsc{xspec} version 12.8.2 (\citealt{Arnaud96})
with an absorbed power law model.  All parameters were left free and
the modified version of the C-statistic available in \textsc{xspec}
was used to determine the best-fit values (\citealt{Cash79};
\citealt{Wachter79}).  Comparing the observation in 2000 with those in
2010 to 2013, there has been modest variation in the best-fit
photon index and the absorption parameter, $n_{\mathrm{H}}$ (Table
\ref{tab:obs}), which was consistently above the Galactic value of
$0.019\times10^{22}\psqcm$ (\citealt{Kalberla05}).  The nuclear flux
varies by up to a factor of a few, which is consistent with previous
studies (eg. \citealt{Harris97,Harris09}).  The spectra for obs. IDs
11513 to 14974 were also fitted simultaneously in \textsc{xspec} to
determine the average best-fit model, weighted by the exposure, for
the ChaRT simulation.  The best-fit parameters for this `stacked'
spectrum are given in Table \ref{tab:obs}.

\subsection*{A3 ChaRT simulations}
\label{sec:chart}


From the nuclear spectral model and the source position in mirror
spherical coordinates, the ChaRT program (\citealt{Carter03}) and the
\textsc{marx} software\footnote{See http://space.mit.edu/CXC/MARX/}
were used to simulate the nuclear PSF.  ChaRT provides an interface to
the SAOTrace raytrace code, which is based on the most accurate mirror
model and incorporates the details of the support structures and
baffles (\citealt{Jerius95}).  ChaRT simulations of the PSF core and
inner wings out to $10\asec$ match well with on-axis observations
(\citealt{Jerius02}).  The M87 nucleus was on-axis in each observation
analysed and the nuclear emission is only significant compared to the
cluster emission to $3\asec$ radius.  Therefore the ChaRT simulations
should provide a good reconstruction of the PSF.  Multiple ray-traces
were used and analysed simultaneously to ensure sufficient photons for
a detailed comparison with the observations.  \textsc{marx} version
4.5 was used to project the ChaRT raytraces onto the ACIS-S detector
and account for the effects of the detector response.
Although some differences exist between the \textsc{marx} ACIS
  effective area and that observed, the excellent agreement between
  the subtracted profiles in Fig. \ref{fig:psfprepostsub} shows that
  this is not a significant effect.  \textsc{marx} also includes the
standard effects of the dither motion and residual blur from aspect
reconstruction errors using the DitherBlur parameter.

From the simulated events file, an exposure-corrected image in the
$0.5-7\keV$ energy band was generated with spectral weighting
determined by the nuclear emission model in section A2.
A surface brightness profile for the simulated PSF was extracted from
this image using a series of concentric annuli of $1\asec$ width.  A
surface brightness profile was also generated for each of the M87
observations from an exposure-corrected image weighted by a spectral
model for the cluster emission in the core (section
\ref{sec:clusterspec}).  The surface brightness profiles for the
obs. IDs 11513 to 14974 were then stacked to produce an average
profile.  The short frame integration times of these observations, and
obs. ID 1808, required the use of a subarray to reduce the readout
time and avoid losing data.  The resulting surface brightness profiles
only extended to a radius of $15\asec$ and therefore, despite strong
pileup in the centre, obs. ID 352 was used to trace the cluster gas to
larger radii.  

The surface brightness profiles for the M87 observations and the PSF
simulations are shown in Fig. \ref{fig:fullradrange}.  The nuclear
emission clearly dominates over the extended cluster emission inside a
radius of $2\asec$.  The profiles appear consistent beyond a radius of
$\sim5\asec$ as expected if all variable point sources have been
excluded from the cluster emission.  Although there has been a
significant reduction in the effective area at low energies over time
due to the build up of contaminant (\citealt{Marshall04}), and this has accelerated since 2009, there is still reasonable agreement in the cluster surface brightness for the region of overlap between the datasets
(Fig. \ref{fig:fullradrange} right).  However, within $\sim5\asec$,
variation in the nuclear flux between the observations produces
significant differences.  The nucleus was brighter for the stacked
observations taken from 2010 to 2013 compared to obs. ID 1808, which
was taken in 2000.  Obs. ID 352 has a $3.2\s$ frame integration time
and is therefore strongly piled up to a radius of $3\asec$, which
reduces the flux detected in the $0.5-7\keV$ energy band.

\begin{figure}
\centering
\includegraphics[width=0.98\columnwidth]{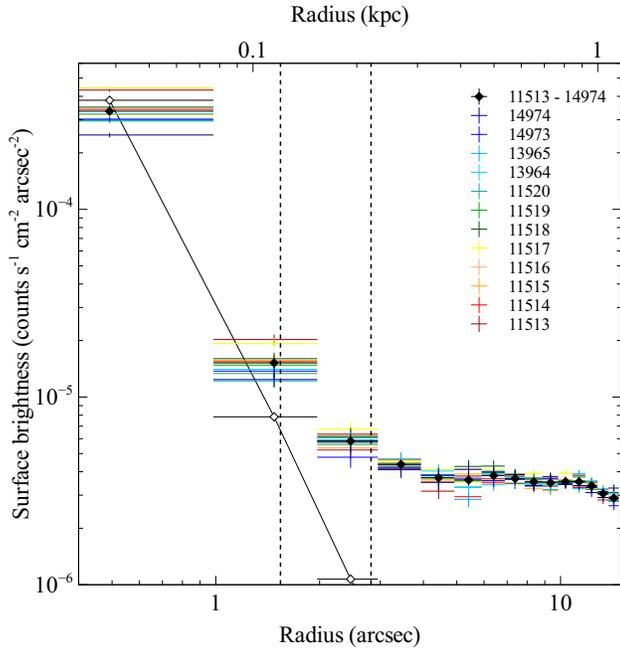}
\caption{Background-subtracted surface brightness profiles in the
  energy band $0.5-7.0\keV$ for the summed and individual obs. IDs
  used in this analysis.  The solid line shows the ChaRT simulation of
  the nuclear flux.  The radial range for the Bondi radius is shown by the vertical dashed lines.}
\label{fig:obsvariation}
\end{figure}



The stacked $11513-14974$ observations were mildly piled up, which
reduced the measured nuclear flux from the central $0-1\asec$ region
over that in the PSF wings.  The PSF simulation was repeated using
\textsc{marx} to include a statistical realisation of pileup
(\citealt{Davis01}).  This piled up simulation was used to estimate
the effect on the nucleus model spectrum and showed that the
$0.5-7\keV$ flux in the central region drops by a factor of $\sim1.15$
with little effect in the surrounding annuli.  The flux of the PSF
simulation was therefore increased to compensate but, while this was a
sufficient correction for the normalization, there was also a low
level hardening of the spectrum.  The soft emission from the hot gas
that is also present at the $\sim5\%$ level within $1\asec$ radius
could reduce this effect.  However, this uncertainty has limited the
temperature measurements at small radii in this analysis.

Fig. \ref{fig:obsvariation} shows the surface brightness profiles for
each of the observations taken from 2010 to 2013.  The nuclear flux
varies modestly over this period but beyond $2\asec$ radius the
profiles are consistent.  We note that the increased uncertainty in
the $5-6\asec$ radial bin is due to the presence of a cavity in the
X-ray emission and there are fewer counts in this region.  Apart from
the central source, there do not appear to be any variable point
sources that have not been removed from the analysis.  The
ChaRT/\textsc{marx} simulations were tested by comparing the
PSF-subtracted surface brightness profiles for obs. ID 1808 and the
stacked 11513 to 14974 observations as described in section
\ref{sec:dataanalysis}.

\bibliographystyle{mnras} 
\bibliography{refs.bib}

\clearpage

\end{document}